\documentclass[aps,prl,twocolumn,floatfix,superscriptaddress,nobalancelastpage]{revtex4}
\usepackage{graphicx}
\usepackage{amsmath,amssymb}
\usepackage{hyperref}
\usepackage{bm}
\usepackage{float}
\usepackage{color}

\begin{document}
\title{Quantum nonlinear phononics route towards nonequilibrium materials engineering: Melting dynamics of a ferrielectric charge density wave}
\date{\today}
\author{M.~Puviani}
\affiliation{Dipartimento di Scienze Fisiche, Informatiche e Matematiche, Universit\`a di Modena e Reggio Emilia, Via Campi 213/A, I-41125 Modena, Italy }
\author{M.~A.~Sentef}
\affiliation{Max Planck Institute for the Structure and Dynamics of Matter, Luruper Chaussee 149, D-22761 Hamburg, Germany}

\begin{abstract}
Negative nonlinear electron-phonon coupling involving an infrared-active phonon mode can lead to an instability towards the formation of a polar lattice distortion with ferrielectric (FE) moments accompanied by an electronic charge-density wave (CDW). Analyzing a small model system in and out of thermal equilibrium, we investigate the FE-CDW and its melting dynamics following an ultrashort laser pulse that drives the ionic dipoles. We observe nonequilibrium coherent phonon amplitude mode oscillations that soften towards the transition to the normal phase. Our case study serves as a first step towards a microscopic understanding of quantum nonlinear phononics as a basis for nonequilibrium control in quantum materials.
\end{abstract}

\maketitle

\section{I. Introduction}

Ultrafast materials science is a blossoming field thanks to new possibilities in driving electronic and lattice excitations to large amplitudes with strong femtosecond laser pulses. Of particular interest are nonequilibrium phase transitions and new light-induced states of matter. Examples for optically steered electronic dynamics include electronic ultrafast photodressing and creation of Floquet states \cite{Wang2013ObservationInsulator,Mahmood2016SelectiveInsulator} with an application in light-induced topological phase transitions \cite{Oka2009PhotovoltaicGraphene,Kitagawa2010,Lindner2011FloquetWells,Sentef2015TheoryGraphene,Hubener2017CreatingMaterials} and the control of electronic screening to modify Hubbard $U$ in strongly correlated materials \cite{Tancogne-Dejean2017UltrafastNiO,Topp2018All-opticalIridates}. 

Mode-specific control of the crystal lattice by direct driving of an infrared-active phonon mode has emerged as an alternative route towards nonequilibrium materials engineering, in particular using the concept of ``nonlinear phononics'' \cite{Forst2011NonlinearControl,Subedi2014a,vonHoegen2018ProbingPhononics}. The underlying key idea is that an infrared-active phonon mode driven resonantly couples nonlinearly through lattice anharmonicities to other normal modes, in particular Raman phonons, in the spirit of ionic Raman scattering. The relevant physics can be understood in terms of classical coupled and driven harmonic oscillators. Examples for resonant control of quantum materials through phononics encompass light-induced metal-insulator transitions \cite{Rini2007ControlExcitation}, the theoretical proposal of phonon-driven Floquet matter \cite{Hubener2018PhononMatterb}, THz-frequency modulation of Hubbard $U$ in an organic material \cite{Singla2015}, or control of magnetism \cite{Forst2015SpatiallyHeterointerface,Nova2017AnDrivenphonons} and superconductivity \cite{Kaiser2014Optically,Mitrano2016PossibleTemperature}. The latter has triggered much theoretical activity \cite{Sentef2016TheorySuperconductivity,Knap2016DynamicalSystemsb,Komnik2016BCSSuperconductivity,Sentef2017TheoryOrders,Murakami2017NonequilibriumDriving}. One particularly stimulating example pertains to the enhancement of electron-phonon interactions in solids, which was observed in bilayer graphene resonantly driven with an infrared-active phonon \cite{Pomarico2017}. The origin of the effect was suggested to possibly lie in a model based on nonlinear electron-phonon coupling \cite{Sentef2017a}, which was independently recognized to open the possibility of light-induced electron-electron attraction via electronic squeezing of pumped phonons \cite{Kennes2017TransientPhonons}. Interestingly, the same nonlinear-coupling model can lead to a coupled electron-lattice instability, similar to the case of quantum paraelectric to ferroelectric transition via electronic doping in SrTiO3 \cite{Edge2015a,Porer2018UltrafastSrTiO3,Wolfle2018SuperconductivitySrTiO}, allowing for ultrafast control of ferroic orders \cite{Buzzi2018ProbingX-rays}, as shown in our study below. We coin the idea of enhancing quantum-mechanical couplings through nonlinear interactions ``quantum nonlinear phononics'' in order to distinguish it from the above-discussed ionic Raman scattering scenario, or ``classical nonlinear phononics'', whose central idea can be explained by classical coupled harmonic oscillators.

Here we investigate a quantum nonlinear phononics route towards phase control in complex quantum materials.  To this end we use exact groundstates and time evolutions of a phononically driven minimal two-site model with quadratic and quartic electron-phonon coupling. We demonstrate that this local model shows precursors of the possible emergence, in the thermodynamic limit, of an intertwined electron-lattice ordered phase with charge-density wave (CDW) order and ferrielectric (FE) ordered moments due to dipoles induced by lattice distortions along an infrared (IR) active phonon coordinate. We further analyze the direct optical control of the ordered phase by a short laser pulse and investigate the coupled lattice and electronic melting dynamics. Light-induced amplitude modes (amplitudons) are tracked and shown to soften towards the phase transition.

\section{II. Model and method}

The equilibrium system is described by a Hamiltonian with a quadratic and quartic coupling of the local electron density to the phonon coordinate:
\begin{align} \label{eq1}
\hat{H}_0 = - J \sum_{\sigma=\uparrow, \downarrow} (\hat{c}_{A, \sigma}^{\dagger} \hat{c}_{B,\sigma} + h.c.) + \Omega \sum_{i=A,B} \hat{b}_i^{\dagger} \hat{b}_i \nonumber \\ 
+ g_2 \sum_{i, \sigma} \hat{n}^{\textit{el}}_{i, \sigma} \left( \hat{b}_i + \hat{b}_i^{\dagger} \right)^2 + g_4 \sum_{i, \sigma} \hat{n}^{\textit{el}}_{i, \sigma} \left( \hat{b}_i + \hat{b}_i^{\dagger} \right)^4 \ .
\end{align}
Here, $J$ is the hopping integral between sites $A$ and $B$, $\hat{c}_{i, \sigma}$ ($\hat{c}_{i, \sigma}^{\dagger}$) the electronic annihilator (creator) operator, $\hat{n}^{\textit{el}}_{i,\sigma} \equiv \hat{c}_{i, \sigma}^{\dagger} \hat{c}_{i, \sigma}$ is the electron number operator, $\hat{b}_{i}$ ($\hat{b}_{i}^{\dagger}$) annihilates (creates) a phonon with frequency $\Omega$ on two sites $i=A,B$. The dimensionless phonon coordinate is expressed as $\hat{x}_i^{\textit{ph}} = \hat{b}_i + \hat{b}_i^{\dagger}$. 
We adopted here the same model Hamiltonian as in \cite{Sentef2017a} with the addition of a further term in the nonlinear electron-phonon coupling: depending on the values of the coupling constants $g_2$ and $g_4$, different materials and phases can be described. We observe that despite being a model with only two sites, the dimensionality of the system is not crucial in this work, since the dynamics is driven by local interactions. This can be thought as the minimal description of a generic cubic perovskite $\text{XYO}_3$, for instance with $X=$ Sr and $Y=$ Ti or $X=$ Li and $Y=$ Nb, with the two sites $A$ and $B$ corresponding to two neighboring unit cells (see Figure \ref{figure1} (a)).

In order to explore the phase diagram with a broken-symmetry state, we consider here only negative values of $g_2$ ($-0.4 \leq g_2 \leq 0$) and positive values of $g_4$ ($ 0 \leq g_4 \leq 0.01$). We fix $J=0.2$ and $\Omega=0.5$, which puts the model into a nonadiabatic regime of fast phonons: the adiabaticity factor as defined in \cite{Fehske2008MetallicityModel} is $\alpha = \Omega / J = 2.5 > 1$. For this reason it is necessary to use \textit{full quantum dynamics} rather than, for example, a quasi-classical Ehrenfest dynamics. In our calculations we used a cutoff of maximally 31 phonons on each site for the bosonic Hilbert space, and convergence in this cutoff was checked.\\

\section{III. Results}

\subsection{A. Equilibrium phases}

\begin{center}
\begin{figure*}
\includegraphics[width=0.7\textwidth]{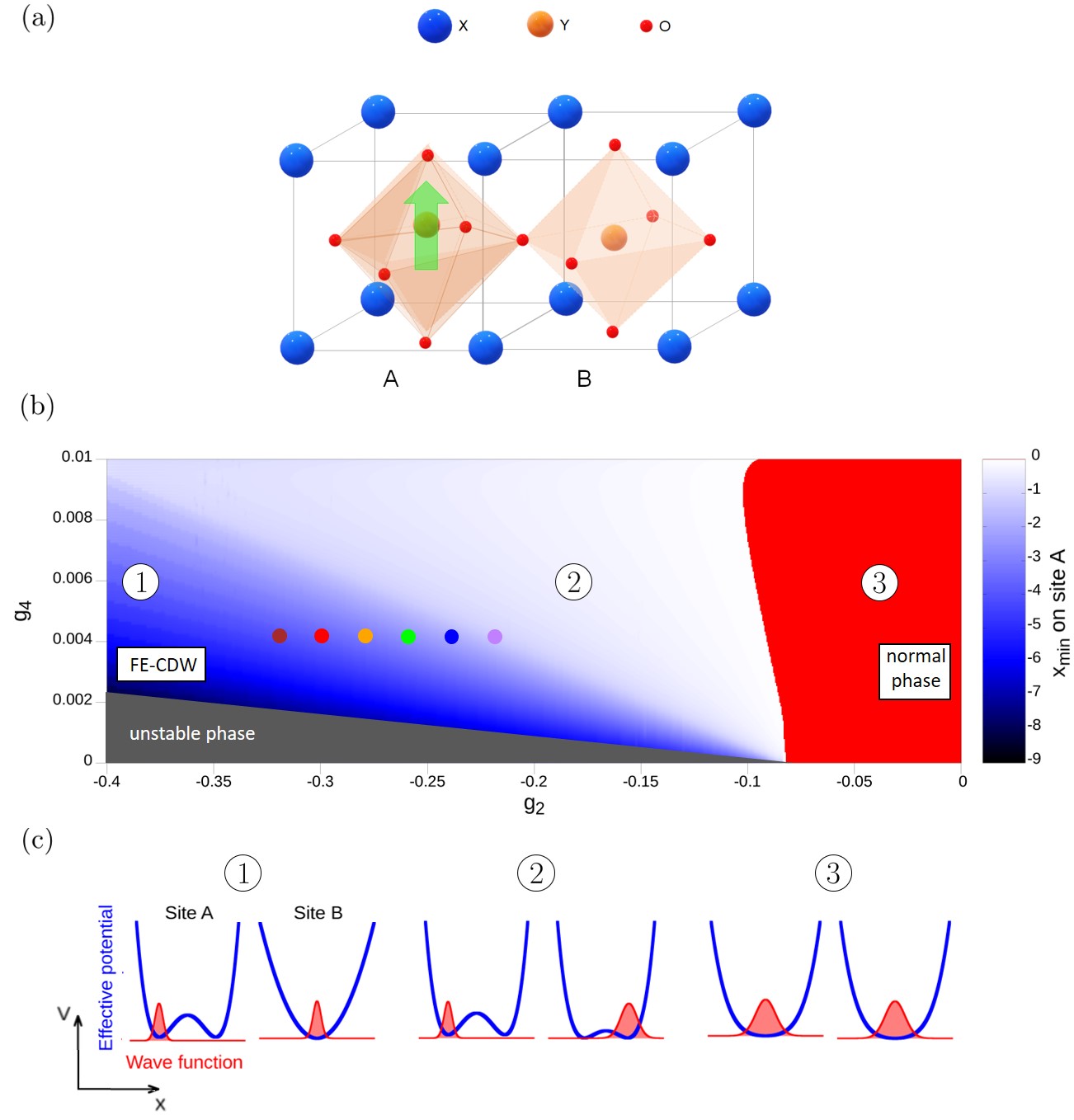}
\caption{\textbf{Equilibrium structure and phases.} (a) Cubic perovskite structure of a generic $\text{XYO}_3$ material with two cells A and B, motivating our minimal two-site model with sites A and B. The ferrielectric phase has the ions displaced along the coordinates of an infrared-active phonon mode at nonzero wave vector, for example with displaced ions in A, resulting in a net electrical polarization (green arrow) for A, and undisplaced ions in B resulting in zero polarization for B. (b) Plot of the phase diagram for the two-site model for different values of quadratic and quartic couplings $g_2$ and $g_4$, respectively. The light and dark blue area is the symmetry-broken (ferrielectric) phase, while the red one corresponds to the homogeneous normal phase. The colored dots correspond to the ($g_2$, $g_4$) values considered in Figures \ref{figure2} and \ref{figure3}. The grey region covers the unstable configuration. (c) Sketches representing the effective phonon potential energy shape (blue) calculated using expression (\ref{eq2}) as a function of the phonon coordinate , and the groundstate phonon wave function (red).}  \label{figure1}
\end{figure*}
\end{center}

In general, coupling nonlinearities are present for any phonon mode but become relevant only at sufficiently large lattice distortions. For the case of ungerade modes in centrosymmetric crystals, the quadratic coupling is the leading one since the bilinear coupling of the ionic displacement (odd) to the electronic density (even) is forbidden by symmetry, because the total Hamiltonian has to be even under spatial inversion. Therefore, our model is geared towards infrared-active phonons. In equilibrium at half-filling (one electron with $\sigma= \uparrow$, one with $\sigma = \downarrow$), we identified three possible phases, as shown in Figs.~\ref{figure1} (b) and \ref{figure1}(c), based on the lattice distortion (electric dipole) and electronic occupation. A normal homogeneous phase (labeled by number 3), occurs for small negative values of $g_2$, or large values of $g_4$. Phase 3 is characterized by a parabolic-like effective potential  on each site. This is physically equivalent to a system with bare optical phonons not coupled to electrons. For more negative values of $g_2$, the effective potential has a double-well shape (phase 2): according to the Ginzburg-Landau semiclassical picture, the most energetically favorable solution breaks spatial inversion symmetry, so that the ground state wave function becomes localized in one of the two energy minima \cite{Zin2006MethodPotential}. These conditions give rise to ferrielectricity (as depicted in Figure \ref{figure1}(c)), since the displacements on each site are typically different in magnitude according to the electronic occupation and with opposite directions (anti-ferrielectric) or same directions (ferrielectric). Eventually, for large negative values of $g_2$, the system acquires a ferrielectric phase (labelled by number 1) with a phononic distortion only on one of the two sites, which also has maximum double-occupation on the distorted site, describing a ferrielectric charge density wave (FE-CDW). For even more extreme and likely unphysical parameters, the model becomes unstable.\\
To gain intuition into the equilibrium behavior, we write a quasi-classical form of the phonon local effective potential, substituting the quantum operator $\hat{x}^{\textit{ph}}$ with the classical variable $x$ for the phonon coordinate, and using the expectation value of the electron number $n^{\textit{el}} = \langle \hat{n}^{\textit{el}} \rangle$, as follows:
\begin{eqnarray} \label{eq2}
V(x) = \dfrac{1}{2} \Omega x^2 + g_2 \ n^{\textit{el}} x^2 + g_4 \ n^{\textit{el}} x^4 \ .
\end{eqnarray}
Analyzing this expression, we distinguish two cases: One with a critical (minimum) point at $x=0$, and another with three distinct critical points, one at $x=0$ (relative maximum) and the others at $x= \pm x_{\textit{min}} \neq 0$ (two symmetric minima). These latter correspond to $x^2_{\textit{min}} = - (\Omega + 2 g_2 \ n^{\textit{el}}) / (4 g_4 \ n^{\textit{el}})$, when $\Omega + 2 g_2 \ n^{\textit{el}} < 0$, assuming $g_4 > 0$ always. We can identify the effective phonon frequency around the equilibrium position, defined as the second derivative of $V(x)$ calculated around the minimum at $x_{\textit{min}}=0$ by $\Omega_{\textit{eff}}= \Omega + 2 g_2 n^{\textit{el}}$. In the second case we calculate it around $x=x_{\textit{min}} \neq 0$, yielding $\Omega_{\textit{eff}}= V''(x_{\textit{min}})$, that is, $\Omega_{\textit{eff}}= -2 (\Omega + 2 g_2 n^{\textit{el}})$. The condition for the FE-CDW to occur is $g_2 < - \Omega / (2 n^{\textit{el}})$. The number of electrons on each site in the groundstate depends on $g_2$ and $g_4$, in addition to the number of phonons on each site (and thus on $x^{\textit{ph}}$). Since the maximum possible occupation on the more highly occupied site is always $1 \leq n^{\textit{el}} \leq 2$, we get in general $g_2 < - \Omega / 4$ as a necessary (but not sufficient) condition to obtain the FE-CDW phase. Vice versa, a sufficient condition is given by $g_2 < - \Omega / 2$.

\subsection{B. FE-CDW melting dynamics}

In order to study the laser-driven dynamics, we set the initial equilibrium condition in a symmetry-broken ground state, either as a FE-CDW or a normal ferrielectric state. We fix $g_4 = 0.004$, and assign to $g_2$ a set of values from $g_2=-0.22$ to $g_2=-0.32$ (pointed as dots in Fig. \ref{figure1} (b)). After preparing the system in its respective groundstate, we apply a short (single-cycle) and strong laser pulse close to resonance with the soft phonon mode. The driven Hamiltonian reads
\begin{align}
\hat{H} (t) = H_0 + F_0 \ e^{- \alpha (t-t_0)^2} \sum_{i=A,B} \sin(\omega t) \left( \hat{b}_i + \hat{b}_i^{\dagger} \right) \ ,
\end{align}
with $\omega$ the laser frequency and $t_0$ the center time of the pulse. We set $F_0 = 1$, $t_0 = 18$, $\frac{1}{\alpha} = 18$ and $\omega = 0.55$ for numerical convenience without loss of generality: in a real experiment, considering materials like strontium titanate, bilayer graphene or organic materials, our parameters choice would lead to a frequency of $\sim$50 THz and a period T$\sim$20 fs \cite{Pomarico2017}.
As in \cite{Sentef2017a}, we coupled the external driving to phonons only, since including also the direct electron-photon interaction would only make the simulation results more complicated without being enlightening for the physics presented here. Moreover, in materials with anisotropic optical conductivity, the selective coupling to phonons only can be obtained with a polarization control of light \cite{Mankowsky2017Dynamical3}.
The initial ground state is evolved using the time-evolution operator as
\begin{align}
| \Psi (t) \rangle = \mathcal{T} e^{-i \int_0^t \hat{H}(t') dt'} | \Psi_0 \rangle .
\end{align}
It was implemented in the calculation code with the commutator-free fourth order scheme as in Ref.~\citep{Sentef2017a}, with a time step $\delta_t = 0.05$ in order to preserve the unitarity of the time-evolution operator and to avoid time discretization issues, up to time $t = 100$.
\begin{figure}
\centering
\includegraphics[width=\columnwidth]{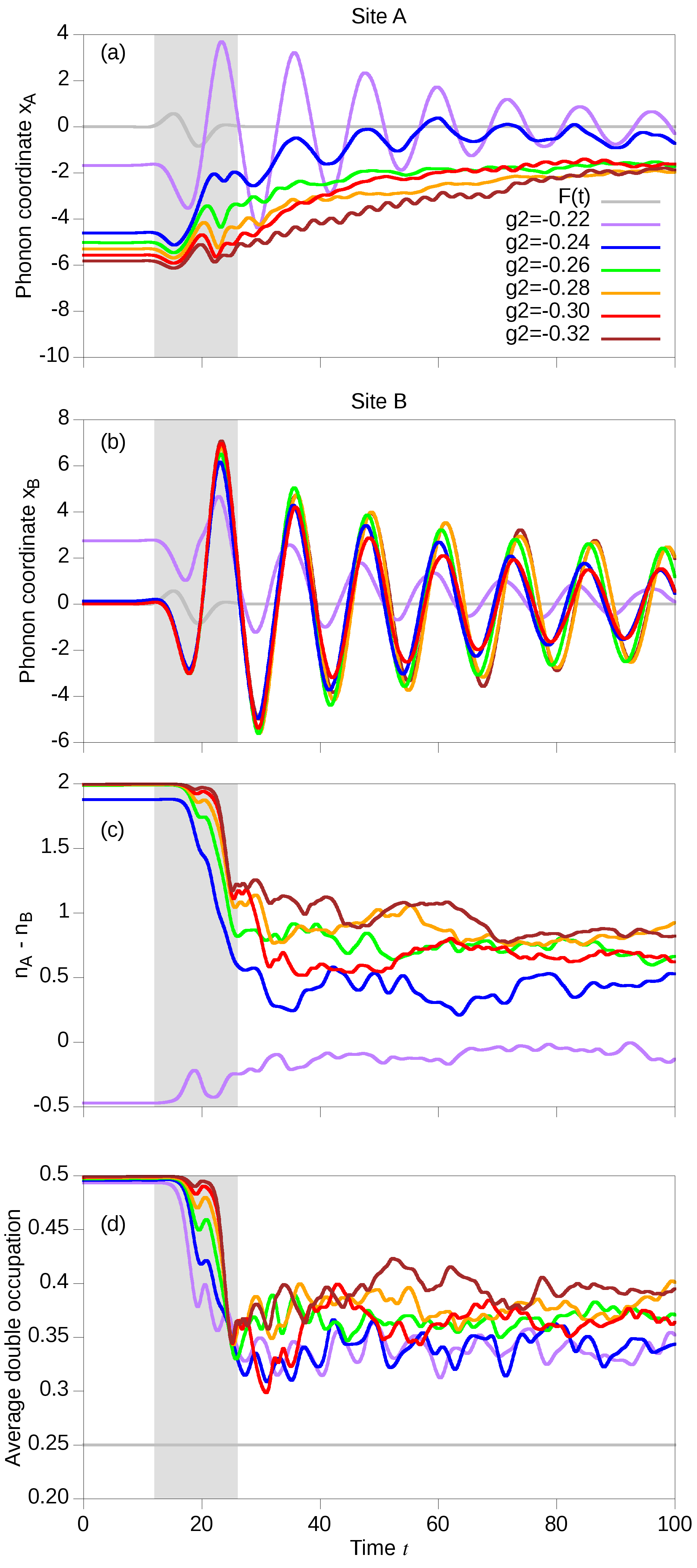}
\caption{\textbf{Light-induced control.} Time-resolved coordinate on site A (a) and site B (b), for different values of $g_2$ as indicated, and $g_4 = 0.004$. (c) Electronic occupation imbalance (CDW order parameter) on the two sites, $n_A - n_B$. The grey line represents the laser pulse $F(t) = e^{-\frac{(t-18)^2}{18}} \ \sin (\omega t)$. The grey area indicates the pulse duration. (d) Electronic double occupation per site. The line at 0.25 represents the random, infinite-temperature value.} \label{figure2}
\end{figure}

When the laser pulse is switched on, the phonons are driven and the coordinates start to oscillate on both sites. On the initially less occupied and less distorted site (site $B$), the ion oscillates around its equilibrium position, and once the laser is turned off the oscillation amplitude is damped [Fig.\ref{figure2}(b)]. The other site ($A$), initially distorted and doubly occupied, is displaced from its equilibrium position when the phonons are driven, and oscillates reducing the structural distortion [Fig. \ref{figure2}(a)]. Due to the anharmonicity of the electron-phonon coupling, the phonon excitations cause a change in the electronic occupations on both sites. This shows up in a dynamical reduction of the double occupancy on site $A$ and an increase on site $B$, at the same time reducing the electronic occupation imbalance thus melting the CDW [Fig.~\ref{figure2}(c)].
The average double occupation decreases during the driving (Fig.~\ref{figure2}(d)). Starting from the fully CDW-ordered value of $\approx$ 0.5, it drops during the laser pulse towards the maximally random, infinite temperature value of 0.25. The reason is that the electronic subsystem is effectively heated by the lattice excitation via electron-phonon coupling. As the FE-CDW is molten, the effective potential that the electrons see is modified in such a way that the electrons on the initially distorted site A tend to move to the initially undistorted site B. Thus electronically delocalized configurations with one electron on site A and one electron on site B become more favorable than in the initial state, leading to the observed decrease in the double occupancy.
Since we are dealing with a closed system, it will not go back to the initial equilibrium ground state. In a more realistic setting, allowing for thermalization processes via a bath coupling, the time scale for the return to the base temperature is typically a few picoseconds \cite{Mitrano2016PossibleTemperature}.\\

\begin{center}
\begin{figure*}
\centering \includegraphics[width=0.85\textwidth]{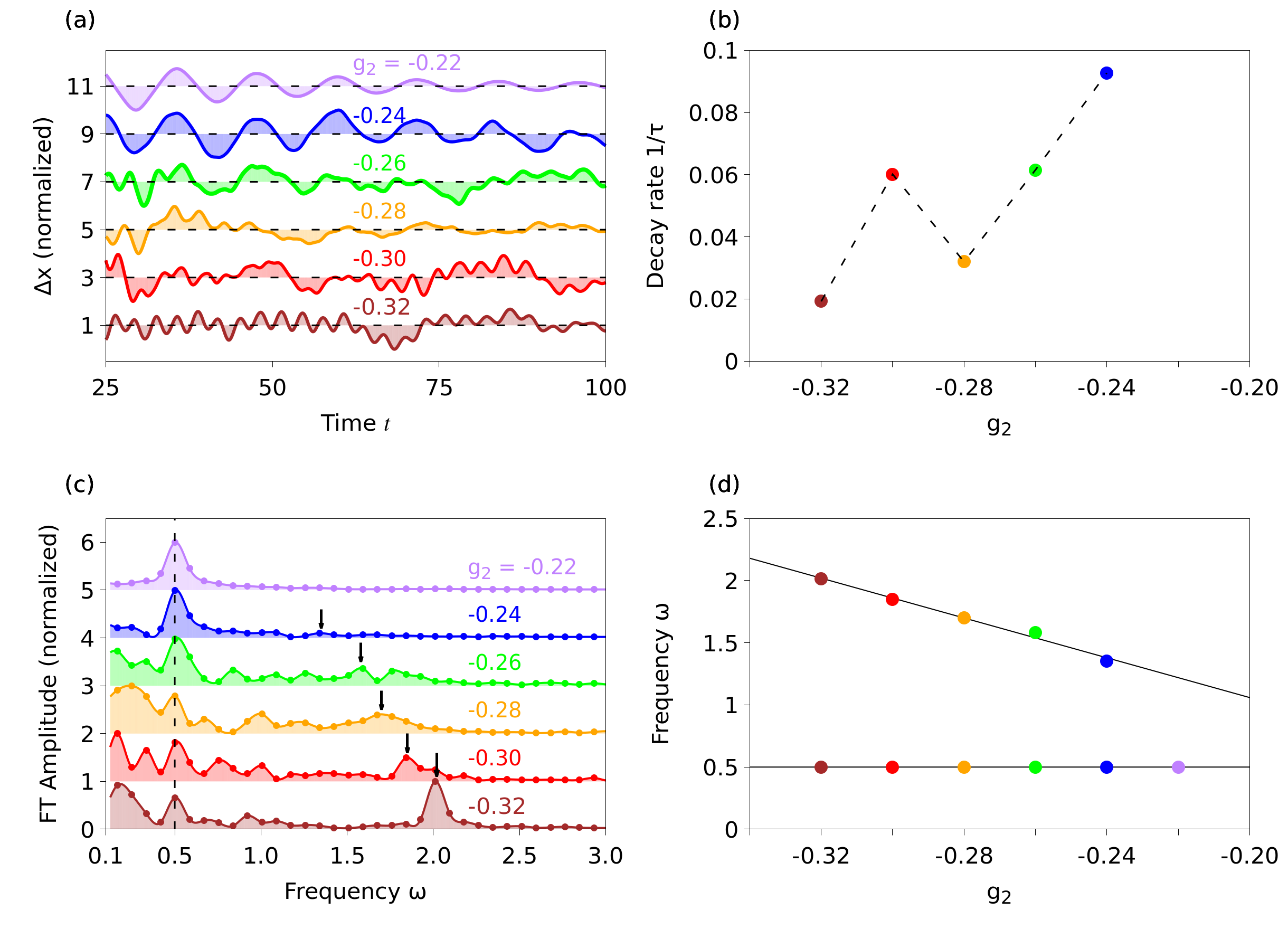}
\caption{\textbf{Amplitude mode oscillations.} (a) Oscillatory motion of the phonon coordinate on site A, after subtracting an exponential fit to the background of the melting dynamics. (b) Decay rate of the phonon coordinate on site A (represented as dots) obtained with an exponential fitting, after subtracting the purely oscillatory behavior. The dashed line is a guide to the eye. (c) Fourier transforms of the oscillatory phonon motion. (d) Dominant peaks (represented as dots) of the oscillations on site A obtained from data in panel (c), highlighting the high-frequency FE-CDW amplitude mode and lower-frequency bare phonon mode.}  \label{figure3}
\end{figure*}
\end{center}

\subsection{C. Amplitude modes}

We now turn to the analysis of the collective amplitude mode oscillations during the melting dynamics: to this aim, we fit the coordinate motion on site $A$ after the laser pulse, starting from time $t=25$, which corresponds to $t-t_0=7$ with respect to the time-center of the driving. We use a fit function of the form $x_A (t) = A \ e^{-t/ \tau} +$ oscillatory terms. We show the purely oscillatory behavior of the phonon coordinate in Fig. \ref{figure3}(a), and the decay rates $1/\tau$ of the melting dynamics after removing the oscillating terms in Fig. \ref{figure3}(b), which has a dip for $g_2= -0.28$. We speculate that this could be indicative of a precursor dynamical critical point, approaching the phase transition between phase 2 and 3 \citep{Tsuji2013NonthermalModel}.
Figure \ref{figure3}(c) shows the Fourier spectrum of the oscillatory motion reported in Figure \ref{figure3}(a). The peak at $\omega = 0.5$ corresponds to the bare phonon frequency and is present in all cases. In addition, there are also higher frequencies peaks for all the initial conditions inside phase 1, i.e. $g2 \leq -0.24$, whose amplitude vanishes when approaching the transition. At the same time, the mode frequency is softened to lower frequencies [Figs. \ref{figure3}(c) and \ref{figure3}(d)]. This is clearly the amplitude mode of the FE-CDW and corresponds to fast oscillations related to the effective phononic potential shape during the melting dynamics. \\
Finally, in order to visualize these complex dynamics,  we investigate how the potential is affected at each time step by the local electronic density and the atomic distortion. To this end we calculate the time-resolved on-site effective potential using the local average electronic density ($\langle \Psi(t) | \hat{n}^{\textit{el}}_i | \Psi(t) \rangle \equiv \langle \hat{n}^{\textit{el}}_i (t) \rangle$) expressed as a function of the semiclassical variable $x_i$ for the phonon displacement, as follows:
\begin{align} \label{Eff_pot}
V_{\textit{eff}} (x_i,t) = g_2 \langle \hat{n}^{\textit{el}}_i(t) \rangle \ x_i^2 + g_4 \langle \hat{n}^{\textit{el}}_i (t) \rangle \ x_i^4 + F(t) \ x_i \ .
\end{align}
The results are reported in \footnote{See Supplemental Material at [URL] for the video showing an example of the potential and lattice dynamics, for $g_2$=-0.28}. This stop-action movie of the lattice dynamics clearly shows that when the laser pulse interacts with the phonons, the potential is stretched in one direction, causing a consequential electronic occupation change, which in turn affects the potential itself. We also note that the quasi-classical potential obtained in this way is not perfectly consistent with the exact dynamics of the quantum phonon coordinate, since the actual dynamics can only be described by a fully quantum process due to nonadiabaticity of the fast phonon.\\ 

\section{IV. Conclusions and outlook}

In conclusion, we have presented a study of a combined lattice and electronic instability in a model with negative nonlinear electron-phonon coupling. Starting from an initial state with intertwined ferrielectric dipole moment and electronic charge-density wave, a short laser pulse resonantly pumping the phonon initiates transient melting dynamics with coherent amplitude mode oscillations that soften and decay in amplitude towards the normal phase. Even though we used a simple model, still a full quantum approach was required to get the right physical insight: the qualitative physical behavior is not affected by the approximations assumed, such as the absence of phonon-phonon interactions, the presence of one only phonon mode, or the lack of coupling to a heat bath. In a broader context, the ultrafast control of phase transitions enabling persistent switching by changing symmetries and functionalities in quantum materials is a field that is still in its development \citep{Stojchevska177}. Our study paves the way for a microscopic theoretical understanding of the interplay of quantum fluctuations of the crystal lattice and electronic configurations, with the infrared-active mode coupling being a particularly exciting one as it can be directly and resonantly accessed by tailored laser pulses. This should be compared and contrasted against similar studies of melting dynamics involving Raman amplitudons in charge-density wave materials \citep{Schmitt1649,Tomeljak2009Dynamics3,Schafer2010DisentanglementSpectroscopy,Liu2013PossibleSpectroscopy,Huber2014CoherentTransition,Schaefer2014CollectiveStudies,Haupt2016UltrafastWave,Singer2016PhotoinducedAmplitude,Mankowsky2017Dynamical3}, or the Anderson pseudospin resonance \cite{Matsunaga1145,Tsuji2015TheorySuperconductors} suggested to be a control knob to launch Higgs modes in superconductors \citep{Matsunaga2012NonequilibriumFilm,Matsunaga2013HiggsExcitation,Mansart2013CouplingSpectroscopy,Kemper2015DirectSuperconductors,Cea2016NonlinearContribution,Krull2016CouplingSuperconductors}. Future work regarding the nonlinear coupling model encompasses an in-depth investigation of the role of adiabatic versus nonadiabatic phonons for the FE-CDW dynamics, systematic studies of larger systems in the thermodynamic limit for example using many-body perturbation theory, or time-dependent density functional theory calculations including first principles estimates of parameter values to understand the role of this type of coupling in materials. \\

\section{Acknowledgments} 
Discussions with A.~Cavalleri, M.~Fechner, and D.~Kennes are gratefully acknowledged. M.A.S. acknowledges financial support by the DFG through the Emmy Noether programme (SE 2558/2-1).
\\
\bibliography{Mendeley_FECDW}

\end{document}